\documentclass[10pt,showkeys,nofootinbib,floatfix,amsmath,amssymb,showpacs,superscriptaddress,notitlepage]{revtex4-2}
\usepackage{hyperref}
\usepackage{caption}
\usepackage{subcaption}
\usepackage{float}
\hypersetup{
colorlinks=true,
urlcolor=blue,
citecolor=blue}
\usepackage[TU]{fontenc}
\usepackage[all]{xy}
\usepackage{amsfonts,amssymb,amsmath,amsgen,amsopn,amsbsy,theorem,graphicx,epsfig}
\usepackage{eufrak,amscd,bezier,latexsym,mathrsfs,enumerate,multirow}
\usepackage{mathtools}
\usepackage{cleveref}
\usepackage[utf8]{inputenc}
\usepackage[english]{babel}
\usepackage[dvipsnames]{xcolor}
\usepackage{academicons}
\usepackage[pagewise]{lineno}
\newcommand{\comment}[1]{}

\begin{document}
\title{Predicting the Masses of Exotic Hadrons with Data Augmentation Using Multilayer Perceptron}
\author{Huseyin Bahtiyar\footnote{Accepted version of International Journal of Modern Physics A}}
\address{Department of Physics, Faculty of Science and Letters, Mimar Sinan Fine Arts University, Bomonti 34380, Istanbul, Turkey}
\date{\today}

\begin{abstract}
Recently, there have been significant developments in neural networks, which led to the frequent use of neural networks in the physics literature. This work is focused on predicting the masses of exotic hadrons, doubly charmed and bottomed baryons using neural networks trained on meson and baryon masses that are determined by experiments. The original data set has been extended using the recently proposed artificial data augmentation methods. We have observed that the neural network’s predictive ability increases with the use of augmented data. The results indicated that data augmentation techniques play an essential role in improving neural network predictions; moreover, neural networks can make reasonable predictions for exotic hadrons, doubly charmed, and doubly bottomed baryons. The results are also comparable to Gaussian Process and Constituent Quark Model.
\end{abstract}

\maketitle

\section{Introduction}	
The mass is a fundamental feature of hadrons, which provides information about the strong interaction. The bare masses of the valence quark (or anti-quark) in hadrons are very small compared to the hadron's effective mass, so most of the mass of matter around us is thought to be a result of strong interaction. The precise measurements help us understand the theories better and develop models. Although theoretical models are quite successful in predicting the masses of exotic particles, results vary between different models. In addition, state-of-the-art lattice QCD calculations require intensive computing power, and in some cases, it is technically difficult to extract excited states or particles near the threshold. Therefore, more studies are necessary to find alternative tools to predict the masses of exotic particles.

The developments in experimental facilities made producing heavy hadrons and composite particles possible. The first exotic particle X(3872) was reported in the $B^\pm \to J / \psi \pi^+ \pi^- K^\pm$ channel by BELLE experiment in 2003 \cite{Belle:2003nnu}. The X(3872) was estimated to contain four quarks with the quark structure $c\bar{u}\bar{c}u$. Investigations of X(3872) have continued and detailed analyzes were made by different experiments \cite{CDF:2006ocq,Belle:2011vlx,BaBar:2010wfc}, until finally in 2013 LHCb collaboration performed an amplitude analysis and quantum numbers were assigned as $J^{PC} = 1^{++}$ \cite{LHCb:2013kgk}. Afterward, more four quark type hadrons have been observed in the charm and bottom quark sectors, such as $Z_c$ \cite{BESIII:2013ris}, and $Z_b$ \cite{Belle:2011aa} respectively.~Recently, pentaquark structures have been observed in LHC \cite{LHCb:2015yax}. 

There is a growing body of literature that recognizes the importance of machine learning and its applications in physics. Neural networks are proven to make successful predictions in nuclear physics \cite{GAZULA19921,ATHANASSOPOULOS2004222,PhysRevLett.124.162502,Yuksel:2021nae,Jiao:2018dro,Jamieson:2022scv}, 
and particle physics \cite{Guest:2018yhq,Radovic:2018dip,Aaij:2012me,Luo:2017ncs,NEXT:2016ire,Psihas:2020pby,Dudko:2020qas,Sombillo:2021ifs}. Also they are useful tools when solving differential equations \cite{PARISI2003715,Mutuk:2020rzm,SUGAWARA2001366}. In lattice QCD, neural networks are mostly employed to reduce the computational cost \cite{Shanahan:2018vcv,Yoon:2018krb,Chen:2021giw}.

It is known that neural networks generally perform better, and their generalizability improve as the training data increases \cite{krizhevsky2012, halevy2009, banko2001}. Predicting masses of hadrons using neural networks is challenging since the number of experimentally observed hadrons is limited; these data are not enough to get high-accuracy results by training neural network models. Considering the deficiency of the available experimental data, applying the data augmentation to the experimentally observed masses of hadrons and investigating the performance of neural networks would be interesting.

Recently, an alternative solution, i.e., data augmentation technique, was developed to improve the results~\cite{Bahtiyar:2022wph}. In Ref~\cite{Bahtiyar:2022wph} experimental uncertainties are used for data augmentation, and the size of the training data set is artificially boosted to determine the root-mean-square error between the model predictions on the testing set and the experimental data. The results show that the data augmentation decreases the prediction errors, stabilizes the model, and prevents overfitting. In light of these developments, this study sought to investigate the usefulness of neural networks in predicting the masses of exotic particles by applying the proposed data augmentation technique.

This paper is organized as follows: Implementation of particle data and detailed information on the neural network architectures are given in \Cref{sec:impandmodel}. In \Cref{sec:res}, results are presented and compared with other works. Finally, the conclusions are given in \Cref{sec:discuss}.

\section{Neural Networks and Experimental Setup}
\label{sec:impandmodel}
The present study aims to build a model that makes predictions of exotic hadrons' masses using their quantum numbers and charges as input. The model takes these properties as an input and predicts the masses as an output. Such problems, where the outcome is a numerical value, are called regression problems, which is a supervised learning problem. In the machine learning terminology, a model is defined as
\begin{equation}
 y= f(x|\phi),
\end{equation}
where $f(.)$ is the model and $\phi$ are its parameters, $y$ corresponds to the prediction for the mass of the hadron. Parameters are optimized by minimizing a loss function. Multilayer perceptron (MLP) is chosen as the model for this work which is an architecture mainly favored for unraveling such regression problems.

Structure of the neural network is built by layers. The first layer is the input layer, and the last layer is the output layer. Layers in between are called hidden layers. Each layer contains a certain number of neurons connected by a weight that provides information about the importance of the neurons. The value of a neuron is calculated by multiplying and summing the inputs $(x_i)$ with the weights $(w_i)$. Then, the weighted sum is inserted into an activation function, a(.), which is generally chosen as a non-linear function. Every hidden layer takes the activation of its preceding layer as input as given in \Cref{eq:firstlayer,eq:secondlayer,eq:generallayer} below. 

\begin{align}
\label{eq:firstlayer}
h^{(1)}_n &= a\left({\textbf{w}^{(1)}_n}^T\,\textbf{x}\right)=a\left(\sum_{i=0}^{d}\, {w^{(1)}_{n\,i}}\,x_i +w^{(1)}_{n_{0}}\right),n=1\dots H_1\\
\label{eq:secondlayer}
h^{(2)}_l &= a\left({\textbf{w}^{(2)}_l}^T\,\textbf{h}^{(1)}\right)= a\left(\sum_{n=0}^{H_1}\,{w^{(2)}_{l\,n}}\,h^{(1)}_n+{w^{(2)}_{l\,n_{0}}}\right),l=1\dots H_2\\
\nonumber
&\dots\\
\label{eq:generallayer}
h^{(m)}_k &= a\left({\textbf{w}^{(m)}_k}^T\textbf{h}^{(m-1)}\right)= a\left(\sum_{n=0}^{H_{k-1}}{w^{(m)}_{k(k-1)}}h^{(m-1)}_{k-1}+{w^{(m)}_{kn_{0}}}\right),k=1\dots H_m\\
\label{eq:outputlayer}
y &= \textbf{v}^T\,\textbf{h}^{(m)}=\sum_{l=1}^{H_m} = v_k h^{(m)}_k + v_0.
\end{align}

Here, $a(.)$ represents the activation function, $\textbf{x}$ is the input layer, the units of the hidden layers are represented as $\textbf{h}^{(1)}_n, \textbf{h}^{(2)}_l$ and $\textbf{h}^{(m)}_k$. $\textbf{w}^{(1)}_n, \textbf{w}^{(2)}_l$ and $\textbf{w}^{(m)}_k$ are the weights belonging to the hidden layers. This calculation propagates in the forward direction. After completing the forward pass, an error is computed using a loss function. The weights are updated in the backward pass using the error. This is how these parameters are optimized during training. The output in \Cref{eq:outputlayer} is computed by taking the last hidden layer activation $\textbf{h}^{(m)}_k$ as input.  In the end, $\textbf{v}$ represents the weights belonging to the output layer. Since the nature of the problem is a regression problem, there is no non-linearity in the output layer.

While designing a model for a supervised machine learning problem, the user is confronted with the tuning of a set of parameters depending on the choice of algorithms or techniques to be employed. At the beginning of the problem, it is unclear which preference to make for these parameters. The selection of the parameters varies according to the problem itself or the given dataset. Parameters that are called hyperparameters, left to the user's choice, vary according to the problem and data set. Choosing the most appropriate hyperparameter group is one of the crucial problems in deep learning. It is essential to correctly build the input layer, especially for physics data with many variables.

In the current study, a total of $432$ meson and baryon masses have been obtained from PDG~\cite{ParticleDataGroup:2020ssz}. Since the number of data is limited, one needs to consider the overfitting carefully. Thus, the input data is randomly divided into two subsets as $90.0\%$ and $10.0\%$ for training and testing, respectively. Since this study aims to estimate the masses of exotic hadrons from the experimental data taken from mesons and baryons, the testing data is partitioned as small as possible. 

Another rule of thumb is to prevent overfitting; it is safer to build a model with more data than the number of parameters \cite{goodfellow2016}. In order to meet this criterion, the number of parameters in the model has to be lower than $432$; therefore, a total of $5$ different architectures, with different hidden layers $16$, $16-2$, $16-4$, $8-8-8$ and $8-8-8-8$ are implemented. RMSProp, Adam, Adam with Nesterov momentum (Nadam), and another Adam-based optimizer Adamax are applied as optimizers. Mean Squared (MSE) and Mean Absolute errors (MAE) are implemented for losses. Glorot Normal and Glorot Uniform are preferred for weight initializers. Finally, for activation, ReLU and tanh functions are chosen. Epochs vary from $1000$ to $5000$ and batch sizes selected as $16, 32, 64$.

\subsection{Preprocessing the particle data}
\label{sub:preprocess}
The particle data has been taken from Particle Data Group (PDG)~\cite{ParticleDataGroup:2020ssz} using python's Particle package~\cite{rodrigues2020scikit}. The data contains the quark structure of hadron, isospin (I), angular momentum (J), and parity (P) quantum numbers, and lastly, charge (Q) and the mass of the particle. The $P$ quantum number is only taken $-1$ or $1$ values. I and J are positive half-integers, and the charge of the hadron is an integer. Some of the particles have the same quark content and quantum numbers but different mass, which would introduce ambiguity. In order to provide unambiguous input to the neural network, another variable called {\em state} (S) is introduced as proposed in Ref.\cite{Gal:2022yqu}. This variable is ordered to rank the particles that share the same quark content and quantum numbers according to their masses using the state value as the distinctive feature for the unclear inputs. In some cases, a particle cannot be identified just by its input features. For instance, without the S variable, a proton and a N(1440) are represented by the same features; however, these correspond to two different particles of different masses. Therefore, the state variable is a key variable for the neural network to be aware of this distinction.

Hadrons contain quarks/anti-quarks with six different flavors. Therefore data can be preprocessed in various ways. In this work, three preprocess types were applied to the particle data gathered from the PDG.

First type of preprocessing is labeling the data set with respect to the quark and anti-quark content of the hadron,  where the values of $q$ and $\bar{q}$ labels are taken to be the bare quark mass in units of MeV. This leads to an input layer with $7$ nodes, 
\begin{align}
\label{eq:NN7}
\text{input }_7&=(q,\bar{q},P,J,I,Q,S)
\end{align}
An example input layer that encodes a positively charged $\pi^+=u\bar{d}$ is,
\begin{align}
\label{eq:pion7}
\pi^+&=(2.16,4.67,0,0,1,1,0)
\end{align}
In the second type of preprocessing, hadrons are labeled by quarks, anti-quarks, and the generation of the quarks. The first generation ($q_1,\bar{q_1}$) contains up and down quarks, the second generation ($q_2,\bar{q_2}$) is composed of strange and charm quarks, and lastly, flavors of the third generation ($q_3,\bar{q_3}$) are bottom and top quarks. These placeholders are also filled with the mass of the specific bare quark. So, the input layer has $11$ nodes,
\begin{align}
\label{eq:NN11}
\text{input }_{11}&=(q_1,\bar{q_1},q_2,\bar{q_2},q_3,\bar{q_3},P,J,I,Q,S)
\end{align}
and an example input layer encoding the positively charged pion into a neural network is,
\begin{align}
\label{eq:pion11}
\pi^+&=(2.16,4.67,0,0,0,0,0,0,1,1,0)
\end{align}
The procedure proposed in Ref.~\cite{Gal:2022yqu} is followed for the last preprocessing type. In this procedure, all quarks and anti-quarks are labeled separately; thus, filling with the bare quark masses becomes redundant. For this reason, the labels are filled according to the number of quarks that make up the hadron in question. Since the top quark does not live long enough to form a hadronic state, $t$ and $\bar{t}$ labels are omitted. Therefore, in this case, the input layer has $15$ nodes
\begin{align}
\label{eq:NN15}
\text{input }_{15}&=(u,\bar{u},d,\bar{d},s,\bar{s},c,\bar{c},b,\bar{b},P,J,I,Q,S)
\end{align}
where a $\pi^+$ state is thus represented as,
\begin{align}
\label{eq:pion15}
\pi^+&=(1,0,0,1,0,0,0,0,0,0,0,0,1,1,0)
\end{align}

\subsection{Data augmentation}
\label{ssec:dataaug}
It is well known that the neural network performs better when the amount of training data increases~\cite{goodfellow2016}. In our case, experimental data on hadron masses are limited. In order to improve the estimations of the model, we employ the prescription given in Ref.~\cite{Bahtiyar:2022wph} to artificially augment our data set.

Data augmentation is frequently used in classification problems of images. In the literature, de-colorization, de-texturization, and flipping/rotating images are main data augmentation methods. Trained with such augmented data, the model makes more successful predictions for images with different positions and orientations~\cite{geron2017}. Since the output of our model is the mass of a hadron (a numerical value), our problem is not a classification but a regression problem. Therefore, one can not directly use the methods listed above. In the literature, adding small random noises to the input is the most frequently used method for regression~\cite{sietsma1991}. Two methods are employed to augment the available particle data.

In the first method~\cite{Bahtiyar:2022wph}, which we call {\it error augmentation}, the experimental errors of masses are added and subtracted to their central values, while the quantum numbers are kept constant. Thus, the training data is resampled twice. For example, experimental data for the mass of $\pi^+$ is $139.57039 \pm 0.00018$ MeV~\cite{ParticleDataGroup:2020ssz}, whereas the new data for $\pi^+$ becomes $139.57039, 139.57057$, and $139.57021$ MeV after resampling. The same procedure is applied for each hadron using their respective experimental errors. Resampling the data for a total of $388$ hadrons, the training data set is increased to $1164$ entries.

Then, the second method, gaussian noise resampling~\cite{Bahtiyar:2022wph} is implemented by using a Gaussian probability density function.
In gaussian noise resampling, random data points are generated using the mean and errors from the training dataset as input, while the quantum numbers of the hadrons are kept constant. The training data is augmented up to $9$ resamplings using this method, resampling the total of $388$ hadrons, the size of the training data set is increased up to $3880$. To demonstrate the procedure, we again consider a $\pi^+$ as an example. Here taking $\mu=139.57039$ and $\sigma= 0.00018$, we resample the $\pi^+$ data using Gaussian probability density function up to $9$ times. We employ a limited number of resampling because one might consider the gaussian noise resampling as a form of smearing. The masses are resampled within their margin of error. Therefore after a certain number of resampling, the results do not improve anymore.

In this work, testing set remains unchanged and only the training set is artificially augmented. Therefore, one can compare the predictions of the models trained with the original, and the augmented data sets and investigate the effect of the data augmentation.

\section{Results and Discussion}
\label{sec:res}
This study aims to predict hadrons the low-lying spectrum of tetraquark candidates, $X(3872)$, $Z(3900)$, $Z(4200)$, $Z(4430)$, $Y(4260)$, $Y(4360)$, $Y(4660)$, $Z_b(100610)$, $Z_b(100650)$ and pentaquark candidates, $P_c(4312), ~P_c(4440), ~ P_c(4457))$ which are recently observed by the LHCb experiment~\cite{LHCb:2019kea}. As a byproduct, we also estimate the masses of the $\Omega_{cc},~\Omega_{cc}^*,~\Xi_{cc}^*,~\Omega_{bb},~\Omega_{bb}^*,~\Xi_{bb},~\Xi_{bb}^*$ baryons predicted by theoretical studies but have not been observed by experiments yet.

\begin{figure}[h!]
\centering
\includegraphics[width=\linewidth]{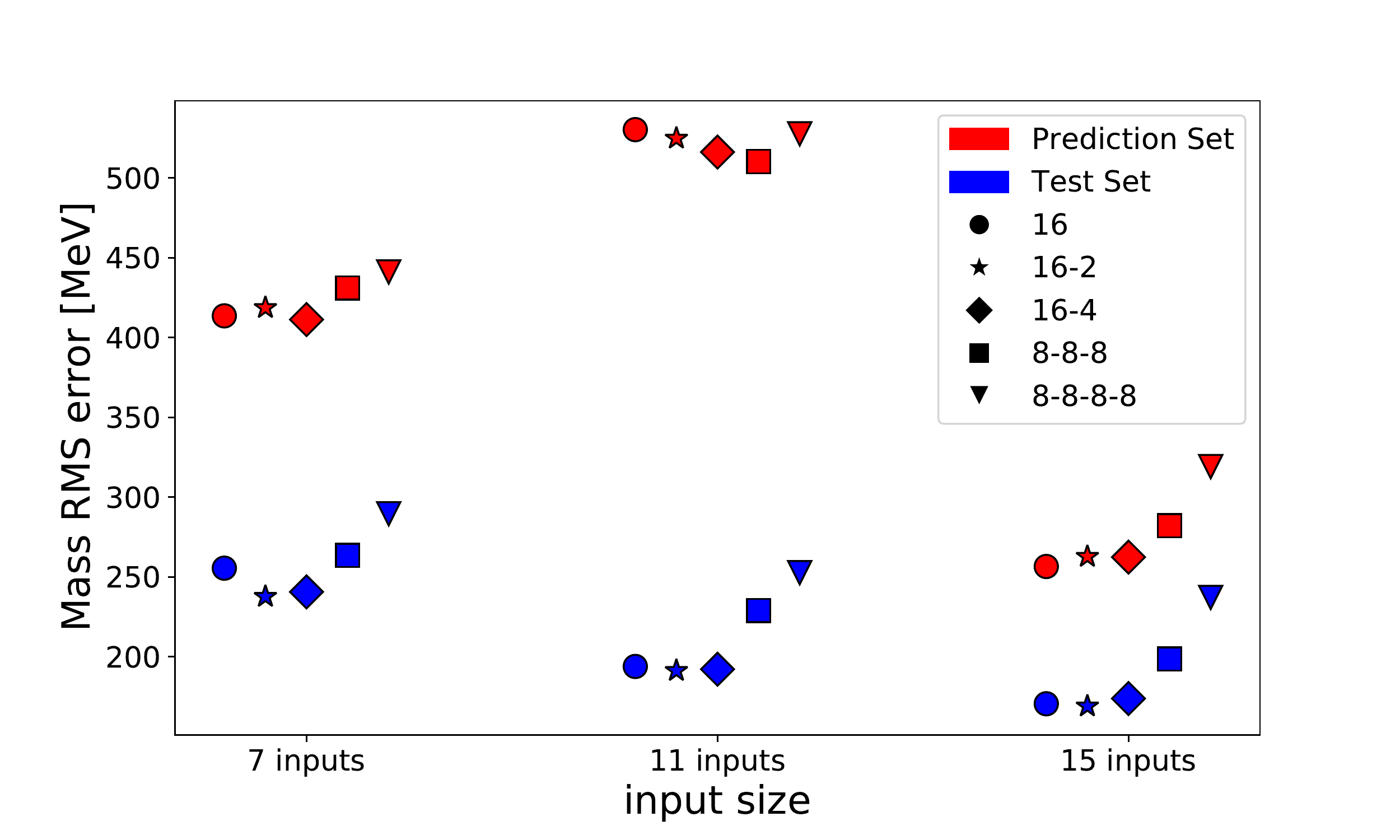}
\caption{The root-mean-square errors of masses for different input sizes. Blue symbols represent errors from testing sets, while the red symbols represent prediction errors of exotic particles. Architectures are categorized by shapes. Hyperparameters are selected as Nadam optimizer, MAE loss, Glorot Uniform weight initializers, and tanh activation function for 4500 epochs and 32 batch size.}
\label{fig:inputs}
\end{figure}

Hyperparameters have been selected as Nadam optimizer, MAE loss, Glorot Uniform weight initializers, and tanh activation function for $4500$ epochs and a batch size of $32$. It is seen from \Cref{fig:inputs}, that the models with seven inputs make the worse predictions from testing data (blue symbols), but when the predictions for exotic particles are examined, the models with 11 inputs give the worse predictions (red symbols). Therefore we have examined the losses of the training and testing sets for possible overtraining for the data of $11$ inputs, and there was no minor gap between training and testing losses. Thus, one can safely conclude that 11-inputs models have made good testing but poor exotic predictions. Finally, \Cref{fig:inputs} reveals that the 15-inputs data set suggested by Ref.\cite{Gal:2022yqu} performs the best both for the exotic and the testing set. In all of the $5$ architectures that have been experimented with, $16,~16-2,~16-4$ architectures gave reasonably close results. For $15$ inputs, RMS errors of the testing set for $16,~16-2,~16-4$ are found as $170.489,~168.979,~173.763$ in units of MeV, respectively, while the RMS errors for prediction (exotics) set are found as $256.519,~262.721,~262.369$ in units of MeV, respectively. In order to minimize the fluctuations coming from randomness, experiments have been performed, and averages have been taken using $10$ different seeds. Due to its success in the prediction set, i.e. having the lowest RMS error, the single-layer model ($16$) has been chosen.

\begin{figure}[h!]
\centering
\includegraphics[width=\linewidth]{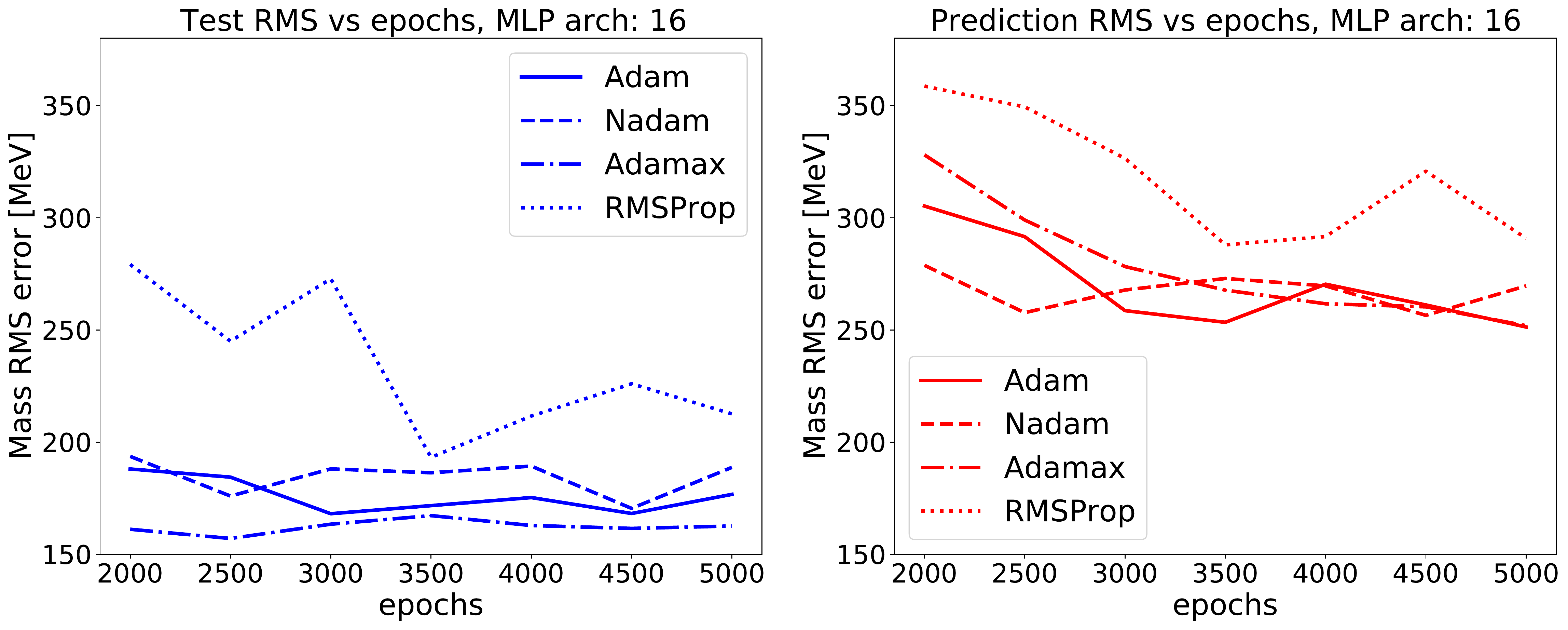}
\caption{The root-mean-square errors of masses for various epochs and optimizers. Blue lines represent errors from testing sets, while the red lines represent prediction errors of exotic particles. Optimizers are categorized by different linestyles. The comparison is performed for a single hidden layer 16-node neural network. Remaining hyperparameters are selected as MAE loss, Glorot Uniform weight initializers, and tanh activation function for 32 batch size.}
\label{fig:optimizers}
\end{figure}

After determining the input size and the architecture, optimal values of the remaining hyperparameters have been determined. Various optimizers have been experimented with and represented in \Cref{fig:optimizers}. Adam, Nadam, and Adamax optimizers perform similarly, while the RMSProp performs poorly compared to others for both training and testing sets.
 
\begin{figure}[h]
\centering
\includegraphics[width=0.5\linewidth]{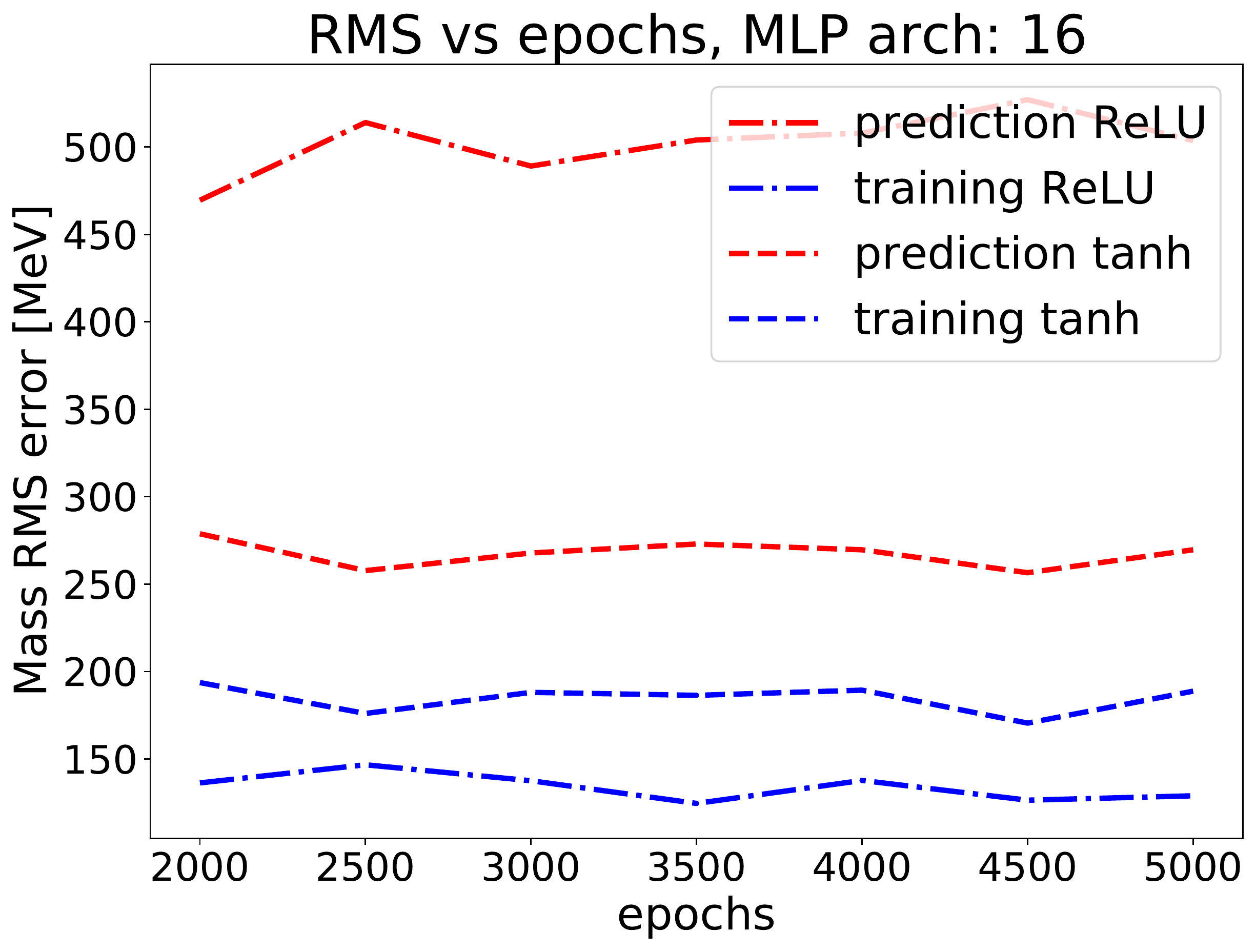}
\caption{The root-mean-square errors of masses for various epochs and activation. Blue lines represent errors from testing sets, while the red lines represent prediction errors of exotic particles. Activations are categorized by different linestyles. The comparison is performed for a single hidden layer 16-node neural network. Remaining hyperparameters are selected as MAE loss, Nadam optimizer, Glorot Uniform weight initializers, and 32 batch size.}
\label{fig:acti}
\end{figure}

A similar comparison has been carried out for the activation functions. Two different activation functions (ReLU and tanh) have been experimented with, and results are shown in \Cref{fig:acti}. The figure indicates that while the ReLU activation function gives better outcomes for the testing set as compared to tanh, the predictions for the testing set are worse than tanh predictions. Since the study's primary purpose is to make a successful mass estimation for exotic hadrons, tanh is chosen for the activation function. As a result, the hyperparameters have been determined as 15 inputs, Nadam optimizer, MAE loss, Glorot Uniform weight initializers, and tanh activation function.

We have augmented the training data using the techniques explained in the \Cref{ssec:dataaug} to examine the changes in the predictions. Using the error augmentation method, an augmented training data set of size $1164$ has been created. Additionally, by applying the Gaussian resampling method, several other training data sets of sizes $n \times 388$ have been created, where $n = 2, 3, \dots, 10$.

\begin{figure}[h]
\centering
\includegraphics[width=\linewidth]{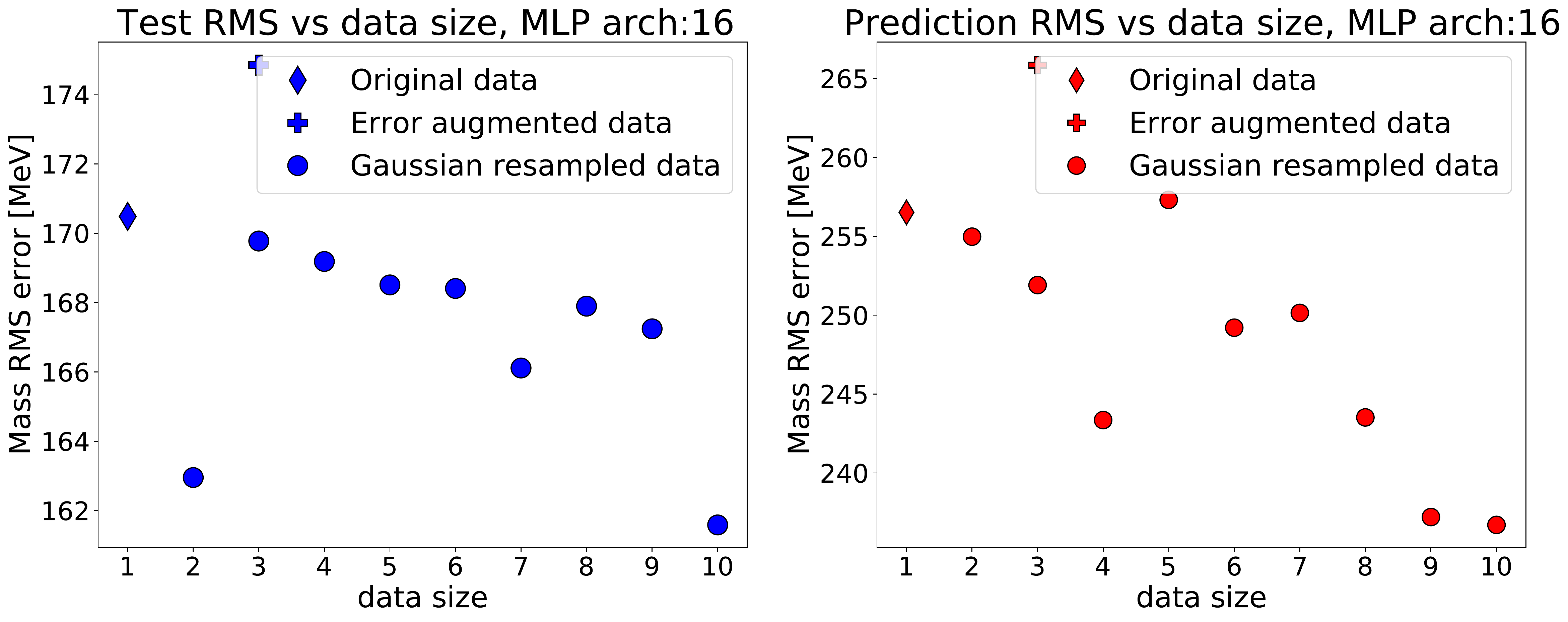}
\caption{The root-mean-square errors of masses for different data sizes. Blue points represent errors from testing sets while the red points represent prediction errors of exotic particles. Thin diamonds represent non-augmented (original) data. Pluses (filled circles) represent error augmented (Gaussian resampled) data (see \Cref{ssec:dataaug} for details). The comparison is performed for single hidden layer 16-node neural network with MAE loss, Nadam optimizer, Glorot Uniform weight initializers, 4500 epochs and 32 batch size.}
\label{fig:res}
\end{figure}

In \Cref{fig:res}, there is a clear trend of decrease in RMS errors when using the Gaussian resampling method. One surprising finding is that the RMS errors are increased for the error augmentation method. This kind of behavior was also reported in Ref.~\cite{Bahtiyar:2022wph}. A slight improvement is observed between RMS errors of original and Gaussian resampled data, around $5\%$ in the testing set and $8\%$ in the prediction set. The observed improvement due to the data augmentation in this work is subpar compared to that reported in Ref.~\cite{Bahtiyar:2022wph}. It is suspected that the reason is the limited size of the training data set (1/3 of that in Ref.~\cite{Bahtiyar:2022wph}), which would limit the efficiency of the data augmentation. Nevertheless, we see a gain by applying the Gaussian resampling method.

\begin{table}[ph]
\caption{Our predictions for the masses of exotic hadrons (in units of MeV) along with a comparison to the Experimental~\cite{ParticleDataGroup:2020ssz} and other Neural Network (NN), Gaussian Process (GP), Constituent Quark Model results~\cite{Gal:2022yqu}.}
{\begin{tabular}{|l|l|l|l|l|l|l|l|} \hline
        Hadron & $I,(J^P)$ &  Structure & This Work & Experiment~\cite{ParticleDataGroup:2020ssz} & NN~\cite{Gal:2022yqu}  & GP \cite{Gal:2022yqu} & CQM \cite{Gal:2022yqu} \\ \hline
&&&&&&&\\[-1em]
        $X(3872)$ & $0\,(1^+)$	&  $c\bar{u}\bar{c}u$ & $3828.19$ & $3871.69\pm 0.17$ & $4815\pm 786$ & $3514\pm {190}$  & $3772$ \\ \hline
&&&&&&&\\[-1em]
        $Z_c(3900)$ & $ 0\,( 1^+)$	& $c\bar{d}\bar{c}u$ & $4075.65$ & $3886.6\pm 2.4$ & \multirow{3}{*}{$4991\pm 815$} & \multirow{3}{*}{$3515\pm {199}$} & \multirow{3}{*}{$3776$} \\ \cline{1-5}
&&&&&&&\\[-1em]
        $Z_c(4020)$ & $ 0\,( 1^+)$ & $c\bar{d}\bar{c}u$ & $4297.87$ & $4024.1\pm 1.9$ &  &  &  \\ \cline{1-5}
&&&&&&&\\[-1em]
        $Z_c(4430)$ & $ 0\,( 1^+)$ & $c\bar{d}\bar{c}u$ & $4520.09$ & $4478^{+15}_{-18}$ &  &  &  \\ \hline
&&&&&&&\\[-1em]
        $Y(4260)$ & $ 0\,( 1^-)$ &  $c\bar{s}s\bar{c}$ & $4049.13$ & $4230\pm 8$  & $5400\pm {1100}$  & $3543\pm {1167}$ & $4072$ \\ \hline
&&&&&&&\\[-1em]
        $Y(4360)$ & $ 0\,( 1^-)$&  $c\bar{u}\bar{c}u$ & $3890.64$ & $4368\pm 13$ & \multirow{2}{*}{$4940\pm 903$} & \multirow{2}{*}{$3107\pm {168}$} &  \multirow{2}{*}{$3772$}  \\ \cline{1-5} 
&&&&&&&\\[-1em]
        $Y(4660)$ & $ 0\,( 1^-)$ &  $u\bar{u}c\bar{c}$ & $4112.86$ & $4643\pm 9$ &  &  &  \\ \hline
&&&&&&&\\[-1em]
        ${\rm P}_c(4312)$ &  $\frac12\,(\frac12^+)$ & $uudc\bar{c}$ & $4171.04$ & $4311.9\pm0.7$& $4200\pm 1200$ & $3544\pm {923}$ & \multirow{3}{*}{$4112$}  \\ \cline{1-7} 
&&&&&&&\\[-1em]
        ${\rm P}_c(4440)$ &  $\frac12\,(\frac12^-)$ & $uudc\bar{c}$ & $4147.95$ & $4440.3\pm1.3$ &  $4100\pm 1100$ & $3253\pm {846}$&\\  \cline{1-7} 
&&&&&&&\\[-1em]
        ${\rm P}_c(4457)$ &  $\frac12\,(\frac32^-)$ & $uudc\bar{c}$ & $4365.89$ & $4457.3\pm0.6$ &  $4500\pm 1100$ & $3581\pm {932}$ & \\ \hline
&&&&&&&\\[-1em]
        $Z_b(100610)$ & $ 0\,( 1^+)$& $b\bar{d}\bar{b}u$ & $10403.09$ & $10607.2\pm2.0$  & \multirow{2}{*}{ $14700\pm 1700 $}& \multirow{2}{*}{$9907\pm {560}$}  & \multirow{2}{*}{$10136$} \\  \cline{1-5}
&&&&&&&\\[-1em]
        $Z_b(100650)$ & $ 0\,( 1^+)$&$b\bar{d}\bar{b}u$ & $10570.77$ & $10652.2\pm1.5$ &  &  &  \\ \hline
    \end{tabular} \label{tab:exotic}}
\end{table}

Recently, a study has been performed to estimate baryon masses from meson masses using the Gaussian Process and neural networks and comparing these results with the Constituent Quark Model~\cite{Gal:2022yqu}. The results from~\cite{Gal:2022yqu} along with our results are listed in~\Cref{tab:exotic}. It is clear that there are significant differences between the masses reported in this study and Ref.~\cite{Gal:2022yqu}. This discrepancy might have been due to the differences in the training datasets, where only mesons data have been used for training in Ref.~\cite{Gal:2022yqu}. Therefore a meson-only dataset might have been sufficient to estimate baryon masses but insufficient for exotics. Consequently, as shown in~\Cref{tab:exotic}, our results indicate that neural networks can make successful predictions for exotics comparable to the Constituent Quark Model.

In addition to the exotic hadrons, predictions have been made for baryons, whose existence is predicted by theoretical models but experimentally have not been observed yet. Predictions have been made for the masses of spin-$\frac12$ $\Omega_{cc}$ and spin-$\frac32$ $\Xi_{cc}^*$, $\Omega_{cc}^*$ baryons in the $c$ quark sector, and $\Xi_{bb}$, $\Omega_{bb}$ and their spin-$\frac32$ partners in the bottomed quark sector. Finally, we have used the neural network to predict the experimentally observed $\Xi_{cc}$ baryon. Predicted masses are given in \Cref{tab:baryon}.

\begin{table}[ph]
\caption{Our predictions for the masses of doubly charmed and bottomed baryons (in units of MeV) along with a comparison to the Experimental (only for $\Xi_{cc}$), Lattice QCD, QCD Sum Rules, and Quark Model results.}
{\begin{tabular}{|l|l|l|l|l|l|l|l|} \hline
        Baryon & $I,(J^P)$&  Structure & This Work & Experiment~\cite{ParticleDataGroup:2020ssz} & Lattice QCD & QCD S. R. \cite{Zhang:2008rt} & Quark Model \cite{Roberts:2007ni}\\ \hline
        $\Xi_{cc}$  & $\frac12,(\frac12^+)$ & $ucc$  & $3673.53$ & $3621 \pm 0.4$ & $3540 - 3680$~\cite{Cheng:2021qpd,Bahtiyar:2020uuj} & $4260 \pm 190$ & $3676$\\ \hline
        $\Xi_{cc}^*$ & $\frac12,(\frac32^+)$ & $ucc$ &  $3891.47$ &---& $3620 - 3750$~\cite{Cheng:2021qpd,Bahtiyar:2020uuj}& $3900 \pm 100$ & $3753 $\\ \hline
        $\Omega_{cc}$ & $0,(\frac12^+)$ & $scc$ & $3713.17$ &---& $3620 - 3760$~\cite{Cheng:2021qpd,Bahtiyar:2020uuj} & $4250 \pm 200$ & $3815$\\ \hline
        $\Omega_{cc}^*$ & $0,(\frac32^+)$ & $scc$ &  $3931.11$ &---& $ 3690 - 3880$~\cite{Cheng:2021qpd,Bahtiyar:2020uuj}& $3810 \pm 60$ & $3876 $\\ \hline
        $\Xi_{bb}$ & $\frac12,(\frac12^+)$ & $ubb$ & $10028.82$ &---& $10143\pm 30\pm 23$~\cite{Brown:2014ena}& $9780 \pm 70$ & $10340$\\ \hline
        $\Xi_{bb}^*$  &  $\frac12,(\frac32^+)$ & $ubb$ & $10243.04$ &---& $10178\pm 30\pm 24$~\cite{Brown:2014ena}& $10350 \pm 80$ & $10367$ \\ \hline
        $\Omega_{bb}$ & $0,(\frac12^+)$ & $sbb$ &  $10067.79$ &---& $10273\pm 27\pm 20$~\cite{Brown:2014ena}& $9850 \pm 70$ & $10454$ \\ \hline
        $\Omega_{bb}^*$ & $0,(\frac32^+)$ & $sbb$ & $10281.95$ &---& $10308\pm 27 \pm 21$~\cite{Brown:2014ena}& $10280 \pm 50$ & $10486$ \\ \hline
    \end{tabular} \label{tab:baryon}}
\end{table}

For the $c$ quark sector, neural network predictions for spin-$\frac32$ particles are around $4\%$ higher than lattice QCD and quark model calculations. For the $\Xi_{cc}$ baryon, the only experimentally observed particle in this table, the discrepancy between the experimental determination and our prediction is only $2\%$. Moreover, our prediction is in good agreement with quark model and lattice QCD results. For the $\Omega_{cc}$ baryon, the neural network's prediction falls within the range of the lattice QCD results and disagrees with other theoretical models. In the bottomed quark sector, our and lattice QCD predictions are in agreement, especially for $\Xi_{bb}$ and $\Xi_{bb}^*$, where our predictions differ by $1\%$. Except for $\Omega_{bb}$, where a $200$ MeV gap is observed between our and lattice QCD predictions. Overall, considering the differences in the predictions between the theoretical models, our data-augmented neural network, which is only a statistical model, have predicted the baryon masses closer to the state-of-the-art lattice QCD results.

\section{Conclusions}
\label{sec:discuss}

The present study is designed to implement neural networks to predict masses of exotic hadrons from ordinary hadrons (mesons and baryons). We use the $432$ experimental data of mesons and baryons from the Particle Data Group as input. Three different input types were used to study the performance of the neural network predictions. It is observed that using categorical variables for quarks in the input layer increases the neural network's performance. The second goal of this study is to investigate the effects of data augmentation on the predictive power of neural network architectures. The experimental error values and the Gaussian augmentation techniques were used to increase the training data set artificially. After selecting the epoch and batch values, we performed ten calculations with different seeds to decrease the fluctuation due to randomization.

The results indicated that data augmentation techniques only improve predictions up to a point. Notwithstanding these limitations, the study suggests that neural networks can make reasonable predictions for exotic hadrons, doubly charmed and bottomed baryons. It is known that neural networks do not include any theoretical physics approach behind, also they are considered as a statistical model. However, our study indicated that it is possible for neural networks to make rapid and reliable predictions allowing their use as an alternative tool.

Further research is needed to explore exotic hadrons using Bayesian neural networks. Moreover, the neural network approaches can be used to train the residues of masses, which could provide valuable insights into understanding the absent physics behind other models.

\section*{Acknowledgments}
The numerical calculations reported in this paper were partially performed at TÜBİTAK ULAKBİM, High Performance and Grid Computing Center (TRUBA resources). The author thanks D. Soydaner, K. U. Can and E. Yüksel for valuable discussions and their comments on the manuscript. The author thanks M.A. Savaş, A. Arslan and F. A. Küçükosmanoğlu for their technical support.

\bibliographystyle{elsarticle-num}
\bibliography{sample}
\end{document}